\documentclass[iop]{emulateapj}

\usepackage{natbib, xspace, color}
\newcommand{\kms}{ km s$^{-1}$\xspace}
\newcommand{\ang}{\mbox{\AA}\xspace}
\newcommand{\nhI}{N$_{\rm H \ I}$\xspace}

\newcommand{\za}{$z_{abs}$}
\newcommand{\lala}{$\lambda\lambda$\xspace}
\newcommand\Lya{Lyman-$\alpha$\xspace}
\newcommand{\dv}{$\Delta v_{90}$}
\newcommand{\OHI}{$\Omega_{\rm H \ I}$\xspace}

\shorttitle{COS Observations of low $z$ DLAs}
\shortauthors{Meiring et al.}

\begin{document}

\title{The First Observations of Low Redshift Damped Lyman-$\alpha$ Systems with the Cosmic Origins Spectrograph\altaffilmark{1}}

\author{J.D. Meiring\altaffilmark{2},
T. M. Tripp\altaffilmark{2}, 
J. X. Prochaska\altaffilmark{3}, 
J. Tumlinson\altaffilmark{4}, 
J. Werk\altaffilmark{3}, 
E. B. Jenkins\altaffilmark{5},
C. Thom\altaffilmark{4}, 
J.M. O'Meara\altaffilmark{6},
K. R. Sembach\altaffilmark{4}
}

\altaffiltext{1}{Based on observations made with the NASA/ESA Hubble Space Telescope, obtained at the Space Telescope Science Institute, which is operated by the Association of Universities for Research in Astronomy, Inc., under NASA contract NAS 5-26555. These observations are associated with program GO11598.}
\altaffiltext{2}{Department of Astronomy, University of Massachusetts, Amherst, MA 01003, USA} 
\altaffiltext{3}{University of California Observatories-Lick Observatory, UC Santa Cruz, CA 95064, USA}
\altaffiltext{4}{Space Telescope Science Institute, 3700 San Martin Drive, Baltimore, MD 21218, USA}
\altaffiltext{5}{Princeton University Observatory, Princeton, NJ 08544, USA}
\altaffiltext{6}{Department of Chemistry and Physics, St. Michaels College, One Winooski Park, Colchester, VT 05439, USA}

\begin{abstract}
We report on the first Cosmic Origins Spectrograph (COS) observations of
damped and sub-damped Lyman-$\alpha$ (DLA)
systems discovered in a new survey of the gaseous halos of low-redshift galaxies.
From observations of 37 sightlines, we have discovered three DLAs and four sub-DLAs. 
We measure the neutral gas density \OHI, and 
redshift density d$\mathcal{N}$/dz,  of DLA and sub-DLA systems  at $z<0.35$. We find 
d$\mathcal{N}$/dz=0.25$^{+0.24}_{-0.14}$ and \OHI=1.4$^{+1.3}_{-0.7}\times10^{-3}$ for DLAs, 
and d$\mathcal{N}$/dz=0.08$^{+0.19}_{-0.06}$ with  \OHI=4.2$^{+9.6}_{-3.5}\times10^{-5}$ for 
sub-DLAs over  a redshift path  $\Delta z=11.9$. 
To demonstrate the scientific potential of such systems, we present a detailed analysis of the DLA
at \za=0.1140 in the spectrum of SDSS J1009+0713. 
Profile fits to the absorption lines determine log N(H I)=$20.68\pm0.10$ with a
metallicity determined from the undepleted element Sulfur  of
[S/H]=$-0.62\pm0.18$. The abundance pattern of this DLA is similar to that
of higher $z$ DLAs, showing mild depletion of the refractory
elements Fe and Ti with [S/Fe]=+0.24$\pm$0.22 and
[S/Ti]=+0.28$\pm$0.15.  Nitrogen is underabundant in this system
with [N/H]=$-1.40\pm0.14$, placing this DLA below the plateau of the 
[N/$\alpha$] measurements in the local Universe at similar metallicities. This DLA has a simple kinematic
structure with only two components required to fit the profiles and
a kinematic width of \dv=52 \kms.  Imaging of the QSO field with HST/WFC3 reveals a spiral
galaxy at very small impact parameter to the QSO and several galaxies within 10$\arcsec$, or 
20 comoving kpc at the redshift of the DLA.  Followup spectra with LRIS on the Keck telescope 
reveal that none of  the nearby galaxies are at the redshift of the DLA. 
The spiral galaxy is identified as the host galaxy of the QSO based on the near perfect alignment 
of the nucleus and disk of the galaxy as well as spectra of an H II region showing emission lines
at the QSO redshift. A small feature appears 0.70$\arcsec$ 
from the nucleus of the QSO after PSF subtraction, 
providing another candidate for the host galaxy of the DLA system. Even with these supporting data, we
are unable to unambiguously identify the host galaxy of the DLA, exemplifying some 
of the difficulties in determining DLA hosts even at low redshift. 

\end{abstract}




\section{Introduction}

Quasar (QSO) absorption line systems provide a unique means to study the
intergalactic medium, as well as the interstellar medium (ISM) of 
galaxies at all redshifts. Resonance absorption lines from metallic ions in the 
ISM of distant galaxies along the line of sight to a QSO provide a powerful
method for the determination of chemical abundances at high redshifts, and also
reveal the properties of low-density gas that is difficult or even impossible to
detect with any other method, even in the nearby Universe. As this method 
is independent of the luminosity of the intervening galaxy, it is a relatively 
unbiased tracer of the chemical evolution of the Universe as a whole. 

\begin{figure*} 
\plotone{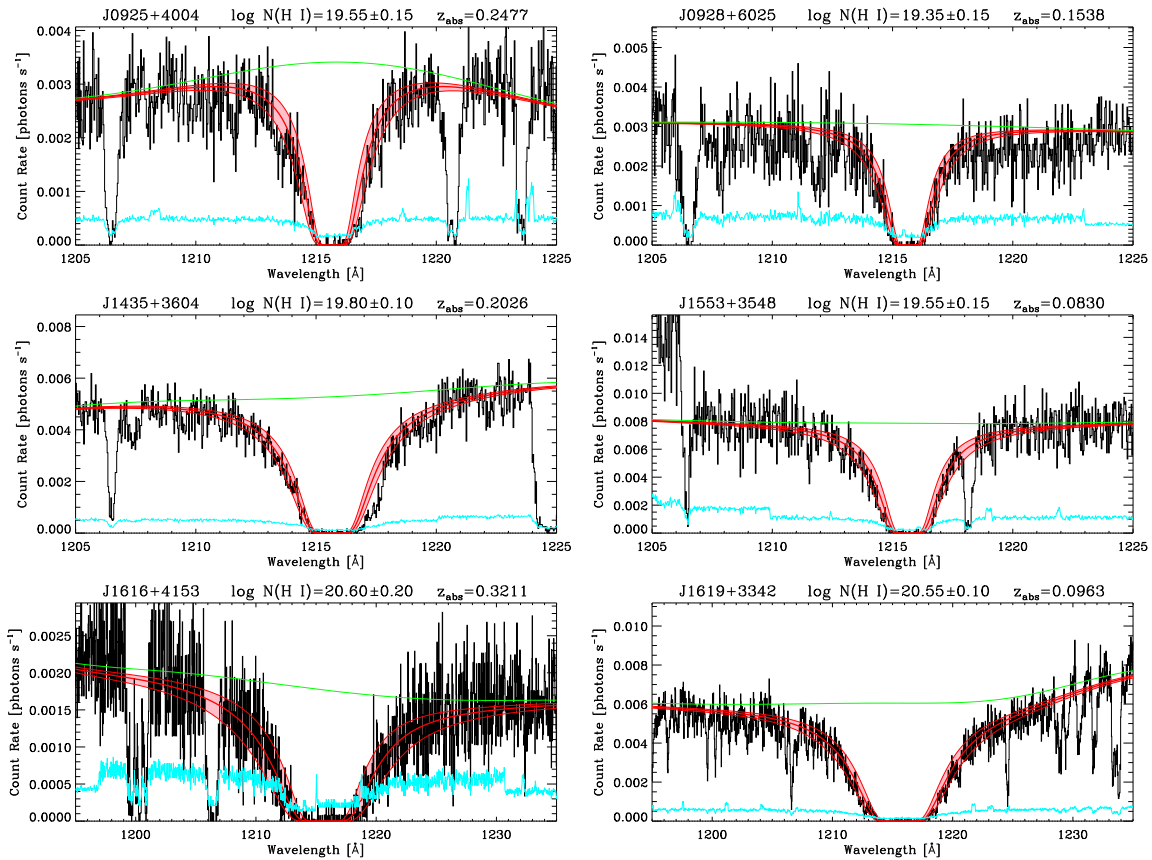}
\caption{ Lyman-$\alpha$ absorption lines from the DLAs and sub-DLAs 
  in this sample. The solid green line is the adopted continuum, and the best fit profiles are
bracketed by profiles with column densities 1$\sigma$ above and below the best fit value. 
The 1$\sigma$ flux uncertainty is shown in cyan below the data.  \label{Fig:Other_DLAs} }
\end{figure*}

The H I $\lambda$ 1215 absorption line in Damped Lyman-$\alpha$ systems (DLAs, log N(H I)$>$20.3) and
sub-DLAs (19.0$<$log N(H I)$<$20.3) can be fit for 
accurate column density measurements due to the extended damping wings in the profile.
 DLAs and sub-DLA systems have been 
shown to contain the majority of the neutral gas in the Universe 
\citep{Wol95, Per03b, Pro05, Not09b}. 

The DLAs are of particular importance for galactic chemical evolution 
studies. With such high column densities, the absorption lines from metal atoms are easily visible. Also, 
systems with high column densities are expected to remain mainly neutral due to self shielding,
 alleviating the need for uncertain ionization corrections to the abundances. 

There is still much debate as to the nature of the galaxies hosting DLAs and sub-DLAs. Even at low 
redshift,  finding a faint galaxy near the bright point source of the QSO is challenging. 
At higher redshift, the cosmological dimming of surface brightness ($\mu\propto(1+z)^4$) makes 
galaxy detections even more difficult. Nonetheless, these systems are a unique laboratory to study the ISM 
of galaxies over a wide range of redshifts.

At $z>1.65$, the Lyman-$\alpha$ line shifts above the atmospheric cutoff at $\sim3000\ang$, and ground based
surveys can efficiently search for DLA systems. Several such surveys have determined the $z>2$ neutral gas density
\OHI \citep{ Wol95, SL00, Pro05, Not09b}. At lower redshifts, space-based spectra are necessary to measure the 
neutral hydrogen column densities in the UV. As DLAs (and QSOs that are bright enough in the 
far ultraviolet (FUV) to be accessible to previous UV spectrographs)  are relatively rare, the number of such 
systems currently known at $z<1$ is small compared to available samples at high redshift.

The Cosmic Origins Spectrograph (COS) is a new instrument package \citep{Fro09}
installed on the Hubble Space Telescope that enables us to study
 these absorbers at low redshift with unprecedented 
efficiency. As $z<0.5$ spans $\sim$ 40 percent of the age of the universe, the low redshift 
absorbers accessible with COS are crucial 
for understanding cosmic chemical evolution and the cosmological gas mass density 
 and for linking their properties to their higher redshift counterparts. 

In this paper, we report on the first DLA systems
 observed with COS. We present a detailed analysis of one
 such system, a DLA in the line of sight to the QSO SDSS J1009+0713, to
illustrate the scientific potential of the observations. 
The structure of the paper is as follows. In $\S$2 we describe in general the HST 
program and data reduction methods we have used for the COS spectra. 
In $\S$ 3 we describe the observations of the field of SDSS J1009+0713, including
the ground based Keck/HIRES spectra of the QSO, the COS UV spectra, and imaging of the field with the 
Wide Field Camera 3 (WFC3). In $\S$ 4 we derive chemical abundances
for this system, discuss the physical state of the gas as determined by the CI and C II* lines, and 
discuss the properties of the galaxies in the field as seen in the WFC3 images. 
In $\S$ 5 we combine the measurements of N(H I) from other absorbers in this program to measure the cosmological
mass density of neutral gas, $\Omega_{\rm H \ I}$, at $z<0.35$ in a blind survey with $\Delta z\sim12$. 
Conclusions are summarized in $\S$ 6. Throughout this paper, we adopt a cosmological 
model with $\Omega_m=0.30$, $\Omega_{\Lambda}=0.70$, and H$_0$=70 km s$^{-1}$ Mpc$^{-1}$.

\begin{figure*} 
\plotone{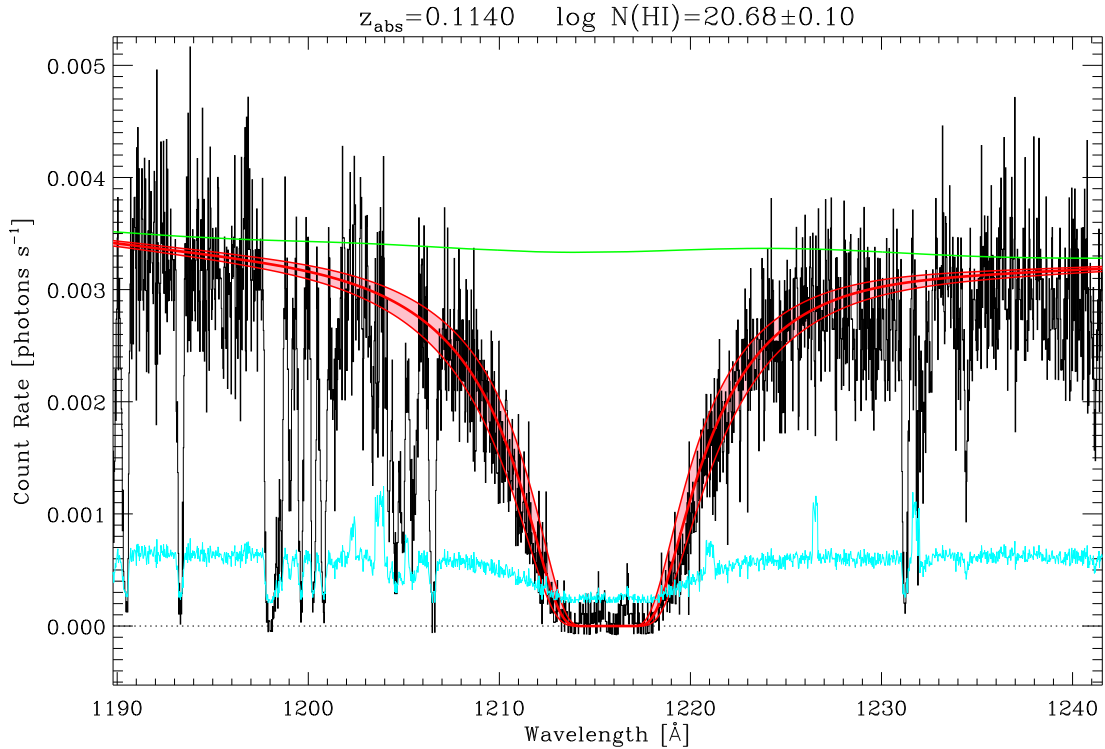}
\caption{ The fit to  the DLA line in the spectrum of SDSS J1009+0713. The solid green line represents
the adopted continuum and the red lines are the best fit model with log N=20.68 bracketed by profiles 
with column densities 1$\sigma$=0.10 dex above and below the best fit value. The 1$\sigma$ flux 
uncertainty is shown in cyan below the data. \label{Fig:DLA} }
\end{figure*}

\section{Program Summary and Observations}

The targets presented here were observed as part of program GO11598, a program
focused on studying multiphase baryons in the halos of $L\ga L^{\star}$ galaxies at 15-150 kpc impact parameters. 
Target QSOs for this program were selected based on sufficient FUV flux and the presence of a 
galaxy seen in the Sloan Digital Sky Survey (SDSS, \citealt{York00}) with impact parameter $\rho<150$ kpc and a spectroscopic 
or photometric redshift $0.15<z<0.35$. Sightlines with strong Mg II
systems seen in the SDSS spectra 
were avoided due to the possibility of a Lyman-limit system (LLS) at $z\la$0.35 which would block the FUV flux. 
In total, 39 sightlines will ultimately be observed in this program. Followup ground-based spectra with the
Low Resolution Imaging Spectrometer (LRIS) on Keck will provide for redshifts, star formation rates and
metallicities of the target galaxies as well as galaxies in the field. 

 These COS data were processed with
\textsc{calcos} version 2.12 during retrieval from the archive.
The individual reduced and wavelength calibrated x1d
  files were then coadded using custom software. 
In short, the gross source and background counts in each wavelength bin are 
  summed after the bins are aligned in wavelength space
by using a cross correlation analysis to allow for small shifts
  due to inaccuracies in the wavelength calibration provided by
  \textsc{calcos}. Simultaneously, an effective integration time at each
  wavelength bin is created which the total counts are ultimately
  divided by, returning the source count rate in photon s$^{-1}$. During
  coaddition, the spectra are background subtracted 
and also corrected for the fixed pattern noise caused by grid wires
above the micro-channel plate detectors using a reference file
provided by the COS instrument team. COS has extremely low backgrounds, 
and consequently, in some cases there are very few counts in the cores of strong absorption lines. 
 We determine uncertainties directly from the accumulated counts, 
using Poisson statistics (see, e.g., \citealt{Geh86}) when the count numbers are low.

The spectra were binned by 3 pixels, as the raw COS
data are oversampled with a $\sim6$ pixel wide resolution
element. All subsequent measurements and analysis were performed on
the binned spectra. The binned spectra have a resolution of $\sim$15 \kms per
resolution element. The S/N of these COS spectra 
 range from $7-15$ per resolution element. 

The DLA and sub-DLA systems that we have discovered in this program 
are shown in Figures  \ref{Fig:Other_DLAs} and \ref{Fig:DLA}. 
The Lyman alpha lines from the systems were fitted using a single component with 
the continuum and profile fit simultaneously.  The redshifts of the systems were determined
 from the metal lines, which with narrow features provide for a more accurate redshift determination. 
In total, we find three DLA systems and four sub-DLA systems in these
COS data. Absorption redshifts and neutral hydrogen column densities for the systems are given in Table 
\ref{Tab:Other_DLAs}. Errors on \nhI given in Table \ref{Tab:Other_DLAs} 
have been estimated by eye and are dominated by the continuum placement uncertainty.

\section{The DLA in SDSS  J1009+0713}

In order to illustrate the capabilities of COS for future observations of DLA systems, we 
present a detailed analysis of one system observed in the sample here. Similar analyses 
for the complete DLA and sub-DLA sample will be given in a forthcoming paper. 
SDSS J1009+0713 ($\alpha$=10:09:02.0, $\delta$=+07:13:43.8, $z_{em}=0.456$), is an
optically bright (m$_g$=17.08,  M$_{g}=-24.7$) QSO observed in the
SDSS. This QSO was chosen for the survey of galaxy halos 
due to the presence of an $L\sim L^{\star}$ galaxy at
an impact parameter of 60 kpc at redshift $z=0.228$.
The DLA described here was serendipitously discovered in the COS spectrum of the QSO.
 No absorption lines from this system are present in the SDSS spectrum of 
the QSO, as the Mg II \lala 2798, 2802 lines which are typically strong enough to be seen in the
SDSS spectra fall at 3115-3125 \ang and are below the wavelength range that the SDSS
spectra cover. In this section, we present a detailed analysis of this system, including the COS and Keck 
High Resolution Spectrometer (HIRES) spectra as well as followup imaging with the Wide Field Camera 3 (WFC3). 

 \subsection{COS Spectra}

These COS data were acquired on  29-30 March 2010 in three orbits as
Visit 13 in program GO11598. Two exposures were obtained with the
G130M grating at central wavelengths of 1291 and 1309 \ang with
exposure times of 1497 and 2191 s. Similarly, two exposures were taken
with the G160M grating at central wavelengths of 1577 and 1600 \ang 
with exposure times of 2002 and 2007 s.

The DLA line of this system was fit with a single component centered at \za=0.1140 with a best solution 
of log N(H I)=20.68$\pm$0.10. The fit to the DLA line is shown in Figure \ref{Fig:DLA}. Profiles with 
column densities 1$\sigma$ above and below the best fit value are also shown in Figure \ref{Fig:DLA}. 

Absorption lines from multiple species are shown in the left panel of
Figure \ref{Fig:Spec}. We detect lines from N I \lala 1134, 1199, 1200,
1201; O I $\lambda$ 1302; Si II \lala  1190, 1304, 1526; S II \lala 1250, 1253, 1259; Fe II \lala 1142, 1143, 1144; Ni II
$\lambda$ 1317. A Lyman limit system (LLS) at \za=0.3556 blocks the
continuum blueward of $\sim$1240 \ang (see \citet{Tum10} for an analysis of the 
LLS) or $\sim$ 1115 $\ang$ in the rest frame of the DLA. 

  There are only two H$_2$ lines covered in the COS spectrum due to the 
presence of the LLS. The Lyman band ($X^1\Sigma_g^+ -  B^1\Sigma_u^+$) transitions 
from the (1-0) band which have rest frame wavelengths $\lambda<1109 \ang$ are all blocked by the LLS. 
Only the H$_2$ \lala 1120, 1125 lines resulting from the (0-0) transitions are covered in
the COS data. These lines are however blended with strong high Lyman series absorption lines
from the LLS; consequently no direct indicator of molecular gas exists for this DLA. 

The COS spectra cover the lines of C IV \lala 1548,1550 as well as the Si IV \lala 1393, 1402 
transitions. The C IV lines lie at the edge of the detector where the sensitivity is low and only the C IV $\lambda$ 1548 line
is minimally detected at S/N$\sim$3. The Si IV $\lambda$ 1393 line is blended with the \Lya absorption line from a 
\za=0.2774 system, and there is no detection of the Si IV $\lambda$ 1402 with S/N$\sim$5 in the region.

 \subsection{Keck-HIRES Spectra}
Optical spectra of SDSS J1009+0713
were obtained with the Keck-II/HIRESb
spectrograph \citep{Vogt94} on the night of 26 March 2010. The S/N of the data ranges from 
$\la$5 per 1.8 \kms pixel  at 3200 \ang to $\sim$20 at 4500 \ang. 
These data provide coverage from 2750-4000 \ang in the rest frame of the \za=0.1140
DLA, and hence include the lines of Mg II \lala 2796, 2803, Mg I $\lambda$ 2852, 
Ti II \lala 3242, 3384, and Ca II \lala 3934, 3969. The Mg I $\lambda$ 2852 line
is partially blended with the Fe II $\lambda$ 2344 line from the \za=0.3558 LLS, as can be seen in Figure
\ref{Fig:Spec}.  In addition, with higher resolution (R=44,000)
than the COS spectra (R$\sim$20,000), these data provide valuable constraints on the kinematics 
and doppler parameters of the lines that are blended at the resolution of COS. 

These data were processed with the Keck/HIRES reduction code \textsc{hiresredux} 
in IDL\footnote{Available at \texttt{http://www.ucolick.org/$\sim$xavier/HIRedux/} }. 
The right panel of Figure \ref{Fig:Spec} shows the HIRES spectra of 
several near-UV (NUV) and optical absorption lines that were observed 
in this DLA. 

The profile was fit with only two components
having doppler widths of $b_{eff}=10.6$ and 12.9 \kms
and centroid velocities of $-6.6$ and $+21.5$ \kms respectively. 
The majority of the low ionization gas appears to be in the $v=-6.6$ \kms component, 
as this component is strongest in all of the non-saturated absorption lines detected here.

\begin{table}
\center
\caption{Properties of the $z_{abs}<0.35$ DLAs and sub-DLAs in this survey. \label{Tab:Other_DLAs}. }
\begin{tabular}{ccccccccccccccc}
\hline\hline 
 QSO                     &             $z_{em}$             &     \za         &           log  N(H I)                      \\
                              &                                      &                    &            cm$^{-2}$                    \\
\hline
SDSS J092554.70+400414.1          &          0.471                  &    0.2477       &     19.55$\pm$0.15               \\
SDSS J092837.98+602521.0         &           0.295                 &    0.1538      &      19.35$\pm$0.15              \\
SDSS J100902.06+071343.8         &           0.456                 &     0.1140     &      20.68$\pm$0.10              \\         
SDSS J143511.53+360437.2          &            0.429                &     0.2026      &      19.80$\pm$0.10               \\
SDSS J155304.92+354828.6          &            0.722                &     0.0830     &      19.55$\pm$0.15               \\
SDSS J161649.42+415416.3          &            0.440                &     0.3211     &      20.60$\pm$0.20                \\
SDSS J161916.54+334238.4          &           0.471              &    0.0963      &      20.55$\pm$0.10               \\
\hline
\end{tabular}
\end{table}

\begin{table}
\center
\caption{AB magnitudes and position offsets for the galaxies near SDSS J1009+0713. Identifiers are based on the position angle and distance from 
the target QSO. \label{Tab:Mags} }
\begin{tabular}{cccccccccc}
\hline\hline
ID           &        $\Delta \alpha$       &      $\Delta \delta$         &           m$_{625W}$           &     m$_{390W}$          &    $z_{gal}$\\
               &               $\arcsec$         &        $\arcsec$                &           AB                         &     AB                       &                                \\
\hline 
170$\_$9          &             +1.58                       &  $-$9.27                          &          21.33$\pm$0.08   &       22.16$\pm$0.12      &     0.35569      \\
86$\_$4            &             +2.87                      &   +0.16                             &           23.21$\pm$0.21   &       24.10$\pm$0.20      &    0.35455        \\
80$\_$5             &             +4.00                       &     +0.81                          &            22.50$\pm$0.12   &      23.09$\pm$0.18      &   0.87899       \\
92$\_$7            &              +6.26                      &     $-$0.44                       &            23.67$\pm$0.37  &      24.45$\pm$0.36        &      1.283?              \\
\hline
\end{tabular}
\end{table}

\subsection{WFC3 Imaging and Keck LRIS Spectra}
  
The SDSS images show no obvious candidates for the \za=0.1140 DLA. In order to determine the nature of
the host galaxy of the DLA system, we acquired imaging in the F390W and F625W filters of this field using the Wide Field 
Camera 3 (WFC3) onboard the HST. Six  exposures were taken in both filters in a dither
pattern to eliminate the effects of bad pixels and cosmic rays. 
Total exposure time was 2370 and 2256 seconds in the F390W and F625W filters respectively.  

\begin{figure} 
\plotone{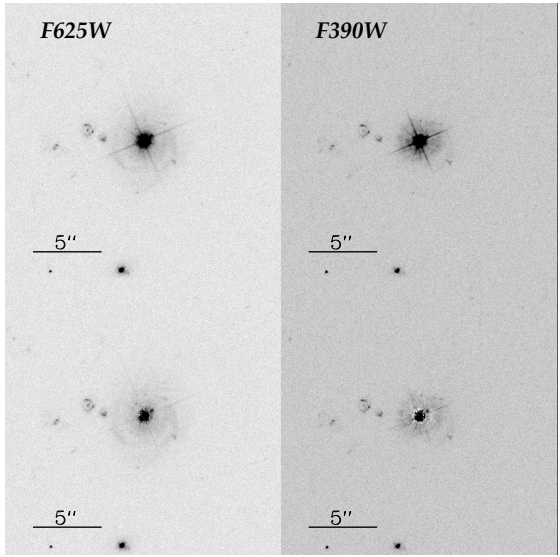}
\caption{F625 and F390W band images of the field of SDSS J1009+0713. The PSF subtracted images 
are shown in the bottom row. \label{Fig:WFC3} }
\end{figure}

\begin{figure} 
\plotone{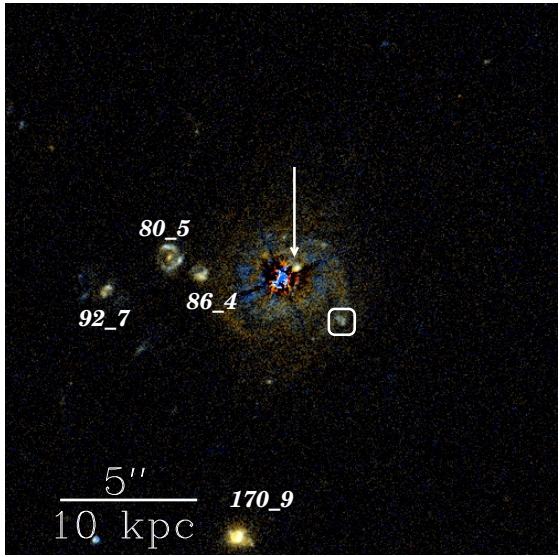}
\caption{Color combined PSF subtracted images of the field of J1009+0713 in the
 F625W and F390W filters. The red channel is the F625W band image,
 the blue channel is the F390W band image, and  the green channel is
 an average of the two. 5 arcseconds is $\sim$ 10 kpc at $z=0.114$. 
The arrow marks the location of the small feature visible after PSF subtraction. An H II region
is marked with a box southwest of the center of the galaxy. North is up and east is
to the left.  \label{Fig:RGB} }
\end{figure}

Even with the superb image quality and resolution that WFC3 enables, galaxies at
small impact parameter to the central bright QSO may only appear after subtracting the QSO with a suitably 
chosen Point Spread Function (PSF). The PSFs used here were acquired from the Tiny Tim  online 
interface\footnote{Available at \texttt{http://www.stecf.org/instruments/TinyTim/tinytimweb} }
at the pixel coordinates appropriate for the position
of the QSO in the images. This PSF was then interactively subtracted from each individual dithered image with
\textsc{idp3}\footnote{Available at \texttt{http://mips.as.arizona.edu/MIPS/IDP3/}}
and the subtracted images were then drizzle combined with \textsc{multidrizzle}. The final drizzled images 
are shown in Figure \ref{Fig:WFC3}. A color combined image was also created from the PSF subtracted images, and 
is shown in Figure \ref{Fig:RGB}. The F625W, F390W, and an average of the two were used in the red, blue, and green 
channels of the color image and the images were scaled using the $asinh$ scaling described in \citet{Lup04}.

Multiple galaxies are seen in these WFC3 images within $10\arcsec$ of the QSO, or 
$\sim 20$ kpc at $z=0.1140$. A large spiral galaxy is clearly seen in Figures \ref{Fig:WFC3} and
\ref{Fig:RGB}. Spectra of objects $80\_5$, $86\_4$, $92\_7$ and $170\_9$ as well as the H II region labelled in Figure \ref{Fig:RGB} were obtained with
 Keck/LRIS\footnote{Labels for galaxies in the field are based on the position angle 
and distance from the QSO; i.e. galaxy 80$\_$5 is at a position angle of 80 degrees 
east of north and 5 arcseconds from the center of the QSO.} .
A 1$\arcsec$ slit was used and aligned such that multiple galaxies and the H II region marked 
in Figure \ref{Fig:RGB} were observed simultaneously. Emission lines from
O III] \lala 4960,5008 and H $\beta$ $\lambda$ 4862 are clearly seen in galaxies $86\_4$ and $170\_9$ placing them 
at $z=0.35458$ and $z=0.35557$ respectively. These two galaxies are likely associated with a LLS at
\za=0.3556 (see \citet{Tum10} for an analysis of the LLS).

\begin{figure} 
\plotone{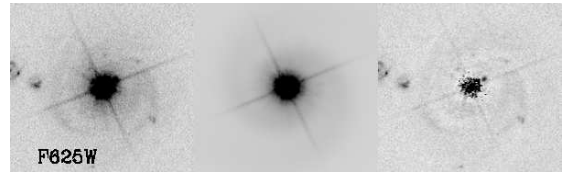}
\caption{Left - The WFC3 image of the QSO. Center - The model adopted from \textsc{galfit}. Right - The residual image. 
Some emission is left over in the point source in  due to saturation effects. All frames have identical scaling. \label{Fig:GalFit} }
\end{figure}

Two emission features from galaxy $80\_5$ 
are seen at 7004 and 8158 \ang, which we interpret as O II] $\lambda$ 3727 and H $\gamma$ 4341 at $z=0.87899$. 
A single strong emission line at 8508 \ang is observed in the spectrum of galaxy $92\_7$. 
No prominent emission lines expected from a galaxy at a redshift $z=0.1140$ 
correspond to this observed wavelength. Assuming this emission line is from O II] \lala 3727,3729 
places this galaxy at $z=1.283$.

Narrow emission lines from  H $\beta$ $\lambda$ 4862 and O III] \lala 4960,5008 from the H II region
 in the spiral galaxy labelled in Figure \ref{Fig:RGB} 
give a redshift of $z=0.45699$. Some flux from the nucleus of the QSO is observed in the spectra, however 
no emission lines from another redshift were seen in the spectra increasing our confidence in the 
determination of the redshift of the H II region. Combined with the close alignment of the nucleus of the QSO
and disk of the galaxy, we interpret the spiral galaxy to be the host of the QSO itself.  

Emission lines from intervening galaxies are sometimes seen
  superimposed on the continuum of the QSO in SDSS spectra \citep{Not10, Bor10}, although most QSOs do
  not show such emission features in their spectrum even when strong absorption
  lines are seen. No narrow emission lines are seen in
  the SDSS spectrum of this QSO at the absorption redshift of the DLA or any other redshift.

\section{Results}

\subsection{Imaging}

We performed aperture photometry on the galaxies in Figure \ref{Fig:RGB} with \textsc{apphot} in \textsc{iraf}. 
AB magnitudes are given in Table \ref{Tab:Mags}. Magnitudes have been corrected for Galactic dust 
extinction\footnote{The online version of the calculator is available at  \texttt{http://nedwww.ipac.caltech.edu/forms/calculator.html}}
from \citet{Sch98}. As galaxies 170$\_$9, 80$\_$5, and 92$\_$7 are
spectroscopically confirmed to be at significantly  higher redshift than 
the absorption redshift of the DLA, an obvious question arises; where is the host galaxy of the DLA?


We have used the \textsc{galfit} software package \citep{Peng10} to fit a multicomponent profile to the F625W image of the QSO. 
A PSF for the nucleus, Sersic profile for the bulge, and exponential disk for the spiral structure was used 
in the fit. The best fit solution from \textsc{galfit} using these components gives m$_B=18.68$, m$_D=18.71$, and
$m_N=17.99$ where $m_B$, $m_D$ and $m_N$ are the magnitudes of the bulge, disk and nucleus respectively. 
The combined magnitude of the bulge and disk is $m_{tot}=17.91$, comparable to that of the nucleus. 
We show the results of the \textsc{galfit} decomposition in Figure \ref{Fig:GalFit}. 

The host galaxies of QSOs are known to be luminous, with L$>$L$^{\star}$, 
and come from a variety of morphological types across the entire Hubble 
sequence \citep{Bah97, Ham02}.  Given a measured apparent magnitude  $m_{\lambda}$, 
the absolute magnitude $M_{\lambda}$ is given by 
\begin{equation}
M_{\lambda}=m_{\lambda}-5\rm{log}\Big(\frac{D_L}{10\ \rm{pc}}\Big) - K_{\lambda} - A_{\lambda}
\end{equation}
where $D_L$ is the luminosity distance, $K_{\lambda}$ is the k-correction and $A_{\lambda}$ is
the extinction due to the Galaxy. Based on the morphology of the galaxy seen in Figure \ref{Fig:RGB}, we assume 
a Sb type spectrum and find a K correction of K$_{\lambda}$=0.25 magnitudes in the F625W filter at $z_{em}=0.456$ and apply an
extinction of $A_{\lambda}$=0.05 magnitudes giving M$_{F625W}=-24.4$. 
The median luminosity of radio loud QSO host galaxies in the sample of \citet{Ham02} is $M_V=-23.5$, although
galaxies classified as having spiral morphologies were  typically more luminous than elliptical counterparts. 

A small feature at small impact parameter appears more clearly in the PSF subtracted images shown in Figure
\ref{Fig:WFC3} and is marked with an arrow in Figure \ref{Fig:RGB}. 
The angular separation from the nucleus of the QSO is 0.7 arcseconds, or $\sim$1.5 kpc at $z=0.114$. 
It is possible that this is the host galaxy of the DLA, or perhaps an H II region 
in the QSO host galaxy. For example, if either of the galaxies 
labelled 80$\_$5 or 86$\_$4 in Figure \ref{Fig:RGB} 
were placed at small impact parameter to the QSO, it would be difficult
to disentangle the two even with the resolution of WFC3. 
The WFC3 grism is likely the best (if 
not only) way to definitively determine the redshift the spiral galaxy and object at small 
impact parameter seen in Figure \ref{Fig:RGB}.

This QSO does show radio continuum emission, with a central source and two lobes detected in the 
FIRST survey  with peak emission of $\sim$ 6 mJy \citep{BWH95}. Radio loud QSOs are almost all found in 
elliptical galaxies; out of the 26 radio loud QSOs observed by \citet{Ham02}, only 4 had spiral type morphologies (16 percent). 

\begin{figure*} 
\plotone{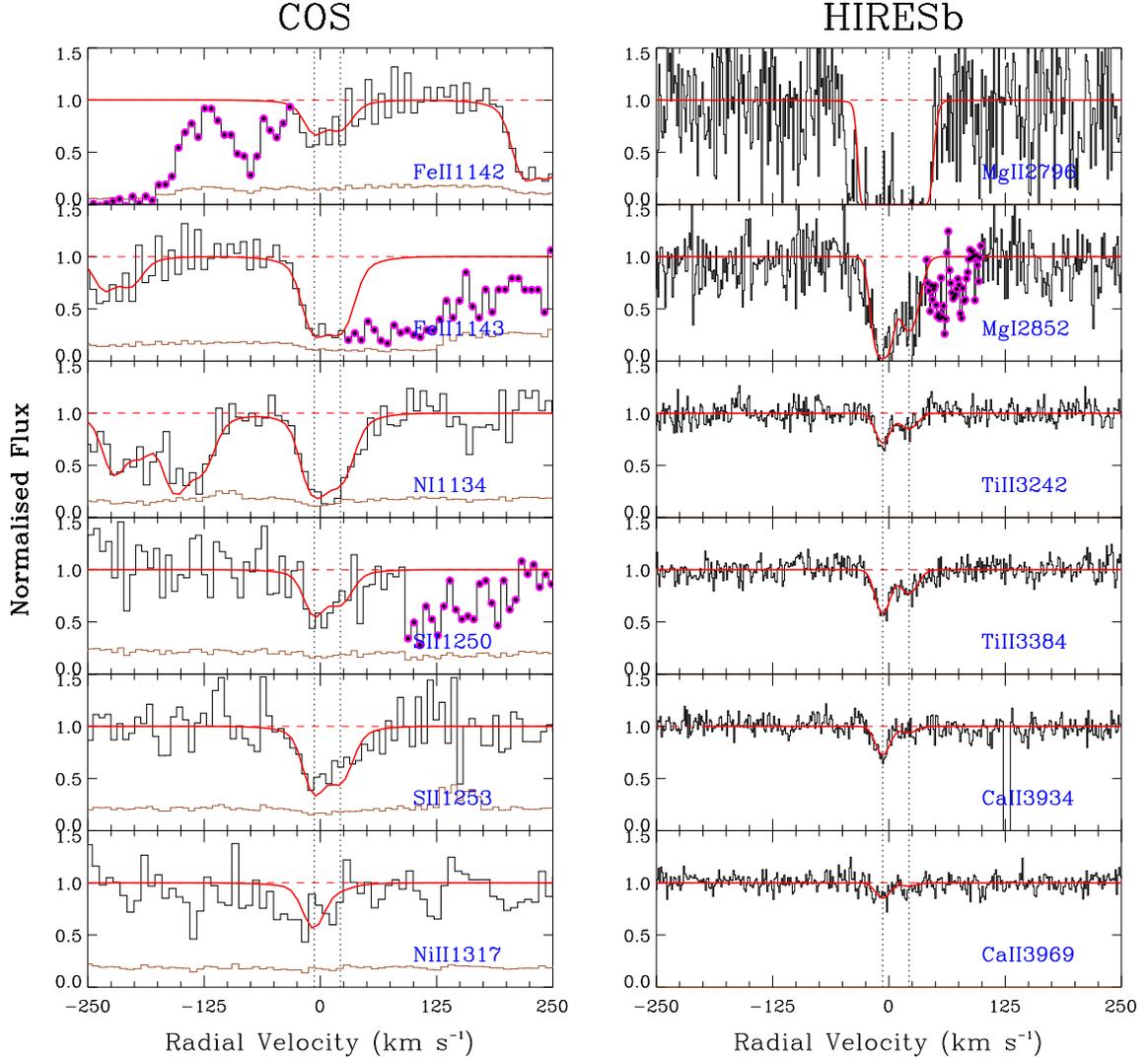}
\caption{Velocity plots for several lines detected in the COS and
  HIRES spectra towards SDSS J1009+0713 at \za=0.1140. The theoretical best fit profile is shown in red, 
and the 1$\sigma$ error in the COS spectra is shown below the data in cyan. The dashed vertical lines denote the positions of 
the centroids of the components used in the fits. Unrelated interloping features from other systems are marked
with magenta circles. Profile fits to nearby members in the multiplets are shown in the N I $\lambda$ 1134 and Fe II $\lambda$ 1142 lines. 
  \label{Fig:Spec} }
\end{figure*}

\subsection{Column Densities and Abundances}
Logarithmic abundances (defined as [X/H]=log X - log H -[X/H]$_{\sun}$)
for the DLA system as determined by the profile fits to the COS and
HIRES data are presented here. Solar system reference abundances ([X/H]$_{\sun}$) are 
taken from \citet{Lodd03}. Rest frame wavelengths as well as $f$ values 
are taken from \citet{Mort03} for all lines aside from the C I lines, for which
we use the revised $f$ values determined in \citet{JT01} and Ni II \lala 1317, 1370 lines
for which we use the  revised $f$ values from \citet{JT06}. 

The COS spectra were fit using the component structure determined with the higher resolution 
ground based Keck spectra. Both the doppler parameters and velocity
centroids were held fixed in the fits, and multiple transitions were 
fit simultaneously  where possible (i.e. the S II \lala 1250, 1253, 1259 lines were all fit simultaneously). 
The synthetic Voigt profiles were convolved with the COS line spread functions 
(LSF)\footnote{Tabulated LSFs for COS are available at \\ \texttt{http://www.stsci.edu/hst/cos/performance/spectral$\_$resolution/}} 
as determined by the COS science team at the nearest tabulated wavelength \citep{Ghav09}. The Keck/HIRES spectra were 
convolved with a gaussian line spread function with FWHM=2.5 pixels. 
Profile fitting was performed using the profile fit code of \citet{Fitz94}. 

  In the case of a particular resonance line not being detected in these data, we have used the 
following procedure to determine a limiting column density for the line. 1) First, we again use the 
doppler parameters and velocity centroids as determined in the higher resolution data. 3) We take the
column densities for one of the better constrained fits to determine the ratio of the column densities 
in each of the components. 3) Synthetic Voigt profiles are then produced and multiplied by  the normalized flux 
and an equivalent width is determined. The integration and statistics are performed over 
a region from the lowest to highest velocity component, with an additional 30 \kms (or $\sim$2 resolution elements) 
on each end to ensure that all of the absorption is taken into account.  4) The significance of the line is determined by
calculating EW/$\sigma_{tot}$ where $\sigma_{tot}=\sqrt{\sigma_c^2+\sigma_p^2}$ and $\sigma_c$ is the uncertainty in the EW due to 
the continuum placement and $\sigma_p$ is the uncertainty This again
perhaps indicates that the spiral seen in 
Figure \ref{Fig:RGB} is not the host of the DLA galaxy. due to photon noise. 5) The column densities are then increased until the
line has a 3$\sigma$ significance. 

  Strong transitions such as O I $\lambda$ 1302, C II $\lambda$ 1334 and Mg II \lala 2793, 2803
are almost universally saturated, and only lower limits to the column densities 
can be  determined by fits to these lines. Lower limits in these cases were
determined by creating synthetic Voigt profiles and increasing the column densities in the components until the 
profile reached the observed flux at the line core.
We have determined a permissible range of  
15.35$<$N(Mg II)$<$16.0 which is equivalent to $-0.9<$[Mg/H]$<-0.2$ 
by measuring an upper limit on the weak transition of Mg II $\lambda$
1239. 

As the component structure in the more highly ionized gas traced by the C IV \lala 1548, 1550 
and Si IV \lala 1393, 1402 lines is often different than that of the low ions and the S/N in the region of the 
C IV lines is too low to discern the component structure, we have measured the column density of C IV
by direct integration of the line using the apparent optical depth (AOD) method. We determine 
log N(C IV)=13.86$\pm$0.20. We also place an upper limit on the column density of Si IV based
on the non detection of the Si IV $\lambda$ 1402 line of log N(Si IV)$<$13.3.

The depletion in this system is mild with [S/Fe]=$+0.24\pm0.22$,
[S/Ti]=$+0.28\pm0.15$ and [S/Ni]=$+0.35\pm0.22$, even less than what is seen in the halo of the Milky Way \citep{Wel97}. 
Depletion levels in the ISM of the Milky Way are seen to vary over two orders of magnitude, from [Zn/Fe]$>+2$ in cold clouds to 
[Zn/Fe]$\sim+0.5$ in the halo of the Milky Way \citep{SS96, Jen09}. We note however that similar mild depletion
was seen at very small impact parameter towards HS1543+593 \citep{Bow05}, a case in which the connection 
between the absorber and host galaxy is unambiguous. 

\begin{table}
\begin{center}
\caption{Total column densities and abundances for the dominant ions in the \za=0.1140 DLA system. \label{Tab:Abund}}
\begin{tabular}{lllllllll}
\hline\hline
Element    &    log N(X)                          &     [X/H]      \\
 \hline
 H            &       20.68$\pm0.10$       &                                                    \\
 C             &        $>$15.90                 &      $>-1.2$                                \\
N             &        15.11$\pm$0.10       &      $-1.40\pm0.14$                   \\
O             &        $>$16.00                  &      $>-1.40$                                \\
Mg           &        $>$15.35, $<$16.0  &       $>-0.9, <-0.2$                               \\
Si             &        $>$15.00                  &      $>-1.2$                                \\
P              &        $<$13.80                    &       $<-0.34$                             \\
S              &      15.25$\pm$0.12        &        $-0.62\pm0.16$                 \\
Ti             &      12.70$\pm$0.03        &        $-0.90\pm0.10$                 \\
Fe            &      15.29$\pm$0.17        &        $-0.86\pm0.20$                 \\
Ni             &     13.93$\pm$0.18        &         $-0.97\pm0.20$                \\
\hline
\end{tabular}
\end{center}
\end{table}

We have used the model of the depletion patterns in the ISM of the Milky Way
from \citet{Jen09} to analyze the depletion in this DLA system.
In this model a single value (F$_{*}$)  indicates the overall level of depletion in a sightline,
with larger values of F$_{*}$ indicating higher levels of dust depletion. 
 We find that the line of sight depletion factor 
F$_{*}=-0.48\pm0.24$ for this DLA, implying an intrinsic metallicity of [Z/H]=$-0.77\pm0.44$, consistent 
with the metallicity determined from the S II lines. 
 Such negative values of F$_{*}$ are typically seen along sightlines with extremely 
low average densities in the halo of our Galaxy \citep{Jen09} implying that the sightline 
passes through the outer edge of a dwarf galaxy. 

Kinematically, the component structure of this DLA is extremely
simple (see Figure \ref{Fig:Spec}). Only two components are required to fit the profiles of the 
absorption lines. The kinematic width \dv (defined as the region where
90 percent of the apparent optical depth in the line occurs) has often
been used as an indicator of the kinematics of QSO absorption line
systems \citep{PW97, Led06}. We find a kinematic width for this system 
of \dv=52 \kms based on the Ti II $\lambda$ 3384 line.

\subsection{Nitrogen}

The nucleosynthetic origins of N seem to include both primary (production of N from newly synthesized C via the CNO cycle)
and secondary production (N produced again in the CNO cycle, but from previously created C and O from previous star formation) processes
in intermediate mass stars \citep{Mar01}. Of the $\sim$30  determinations of N abundances
 in DLA systems, all are at high $z$ where the N I lines
are redshifted into wavelengths accessible with ground based telescopes \citep{Pet08, Hen07}. 
Some constraints have been placed on nitrogen in low-z sub-DLAs where, 
like the high-z DLAs, Nitrogen is found to be underabundant (e.g.,\citet{Tripp05}).

   Abundances from N can be readily obtained from optical spectra of H II regions in nearby star forming galaxies of a range
of luminosities. The low luminosity galaxies from \citet{Zee06} show a flat plateau in the [N/O] ratio with respect to metallicity, 
while the higher metallicity H II regions in \citet{Zee98} show a sharp increase in [N/O] with increasing metallicity, indicative 
of secondary production of N. 

\begin{figure} 
\plotone{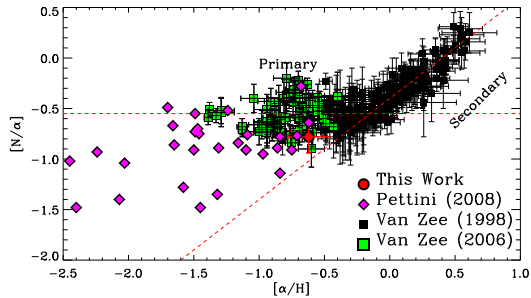}
\caption{ [N/$\alpha$] vs. [$\alpha$/H] for this system (red point),
  high $z$ DLAs from \citet{Pet08}, H II regions in nearby dwarf
  galaxies from \citet{Zee06}, and 
H II regions in nearby massive spirals from \citet{Zee98}. The dotted lines denote the solar abundance. \label{Fig:Nitrogen} }
\end{figure}

   Observations of N in extremely low metallcity DLAs are important for the study of N production and galactic chemical
evolution in general \citep{Pet08, Cent03}. It is still yet to be seen if the [N/O] ratio in DLAs displays the same upturn 
as metallicity increases as most DLAs are metal poor and at high $z$ the N I lines
are often blended with the Lyman-$\alpha$ forest. The DLA presented here is the highest metallicity DLA with both N and $\alpha$ 
abundance measurements. 

  In Figure \ref{Fig:Nitrogen} we show [N/$\alpha$] vs [$\alpha$/H] for H II regions in nearby galaxies from \citet{Zee98, Zee06}, as well as
the DLA sample from \citet{Pet08} and this DLA. The DLA from this work is shown with the red circular point, and lies below  
the ``knee'' of the diagram where N production transitions from primary to secondary production. The mean [N/$\alpha$] ratio 
from the samples of \citet{Zee98} and \citep{Zee06} with [$\alpha$/H]$<$-0.3 is $\langle$[N/$\alpha$]$\rangle=-0.55\pm0.15$,  
3$\sigma$ higher than seen in this DLA of [N/S]=$-0.78\pm0.16$. \citet{Pet08} note this discrepancy in the 
sample of DLAs shown in Figure \ref{Fig:Nitrogen} and propose that the lower values seen in high $z$ DLAs could be due to the 
delayed enrichment of N with respect to O as the delay of $\sim$250 Myr is  a greater fraction of the time available for star formation
at high redshift than in local galaxies, so at high $z$ this effect should be more pronounced. However, in this low-$z$ DLA we see the same
decrement in [N/$\alpha$] compared to the values from local H II regions. 
It would be interesting to see if other high metallcity DLAs are also below the plateau, perhaps indicating a systematic
offset in abundances determined via absorption and emission line diagnostics.

\subsection{C I and C II$^*$}

Both C I and C II$^{*}$ are important diagnostic resonance lines in the 
study of the ISM.  Emission from the 158 $\mu m$ emission line of singly ionized
carbon in the ISM of the Milky Way is the dominant coolant, and is
expected to also  be so in high-$z$ DLAs as well.  This line originates from 
the transition between the $^2P_{3/2}$ and $^2P_{1/2}$ levels in the ground state 
of C$^+$. In the cold neutral medium (CNM, T$\sim10^2$K) the excited state is populated 
largely by collisions with hydrogen atoms, where in the more diffuse warm neutral medium 
(WNM, T$\sim10^4$K) both collisions with free electrons as well as photon
pumping can populate the excited state \citep{WPG03, Bow05}. 

In the WNM the electron density can be determined from observations of the C II and C II$^*$ lines as
\begin{equation}
n(e)=18.3\sqrt{T_4}e^{0.091/T_4} \Big[ \frac{\rm N(CII^*)}{\rm N(CII)} \Big] \ \rm cm^{-3}
\label{Eq:eden}
\end{equation}
where T$_4$ is T/10$^4$K \citep{Leh04}. In the ISM of the Galaxy, the CNM is characterized by having large depletions
of the refractory elements such as Fe and Ti relative to the volatile elements of S and Zn which do not 
condense onto dust grains. As the depletions levels in this DLA are modest, 
with [Ti/S]=$-0.37\pm0.12$, we assume that the sightline is passing through mainly a WNM. 
Depletion levels seen in the CNM of the Milky Way are typically ten times higher than what is 
observed in this DLA \citep{SS96, Jen09}. 

\begin{figure} 
\plotone{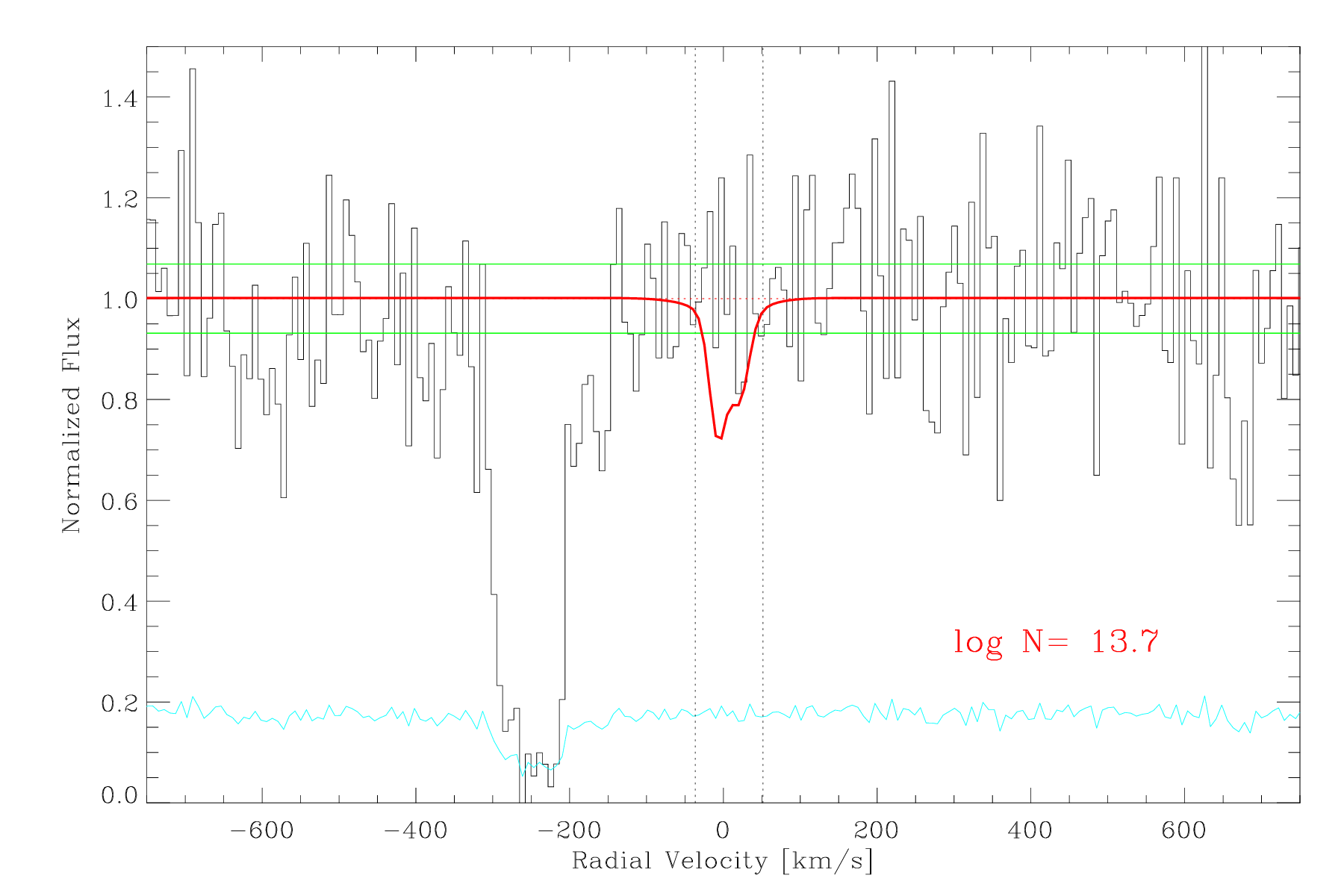}
\caption{ Synthetic Voigt profiles of the CII* \lala 1335.6, 1335.7 lines overplotted on the COS data. We estimate the 
limiting column density of CII* to be  log N(C II*)$\sim13.70$.   \label{Fig:CIISTAR} }
\end{figure}

 The C II$^*$ $\lambda$ 1335 line was not detected in these data, with S/N$\sim$11 per resolution element
in the region around the line. We have used the procedure discussed above to determine a limiting column density
of  log N(C II$^*$)$<$13.7. We  show a synthetic Voigt profile with this total column density overplotted on the
data in Figure \ref{Fig:CIISTAR}.

As was mentioned above the C II $\lambda$ 1334 is strongly saturated and only a limiting column density 
of log N(C II)$>$15.90 can be determined from this transition. However, in sightlines with low depletion such as 
this Carbon is mildly depleted and we can use the S abundance to estimate a total C II
column density  assuming [C/S]$\sim$0.0. 
The above analysis yields log N(C II)=16.45. Using the above values in Equation \ref{Eq:eden} 
with a temperature  of 6000 K characteristic of the WNM the electron density 
is determined to be n(e)$<$0.08 cm$^{-3}$.

The cooling rate per H atom due to C II] $\lambda$ 158 $\mu m$ emission can be estimated by 
\begin{equation}
I_{c}=2.89\times10^{-20}\Big[\frac{ \rm N(C\ II^*)} {\rm N(H\ I)}\Big]  \  \rm{erg\ s^{-1}\ (H\ atom)^{-1}}
\label{Eq:emis}
\end{equation}
\citep{Leh04}. Using the values above, we determine log $I_c<-26.5$. This is typical of what has
 been observed in higher-$z$ DLA systems \citep{WPG03, Leh04}. 

  C I is a tracer of cold gas in the ISM, and is usually seen in DLA absorbers bearing H$_2$ \citep{Sri05}. 
As such, C I absorption lines are indicative of cold, dense material in DLAs. 
The C I $\lambda$ 1277 line provides the strongest constraint on the C$^0$ 
column density in the COS spectrum as the S/N in the region around the 
stronger transition of C I $\lambda$ 1560 was much lower. This line was not detected in the spectrum and we have 
determined a limiting column density of log N(C I)$<$13.6 using the prescription above. 
In the two phase ISM model of \citet{Liszt02}, the ratio of N(C II)/N(C I) is indicative of the gas
density. The observed ratio of N(C II)/N(C I)$>$700 in this system implies 
that n(H)$<$10 cm$^{-3}$.

\begin{figure} 
\plotone{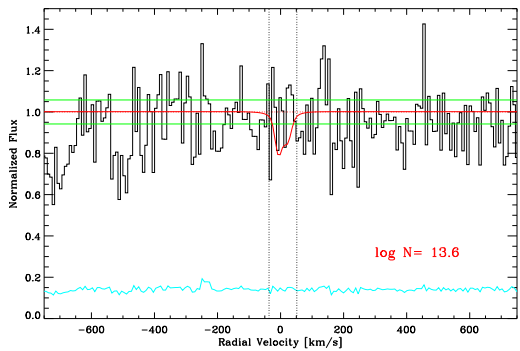}
\caption{ Synthetic Voigt profiles of the C I $\lambda$ 1277 line overplotted on the COS data. We estimate the 
limiting column density of CII* to be  log N(C II*)$\sim13.6$.   \label{Fig:CI} }
\end{figure}


\section{The Cosmological Neutral Gas Mass Density}
The sample of QSO sightlines from this HST program provides an opportunity to 
determine the incidence of DLA systems at low $z$, where there is a paucity of data
due to the necessity of space-based observations. Ideally, a totally ``blind'' survey 
in which there is no prior information about the 
sightlines prior to observation provides for the least amount of systematic biases
in the determination of cosmological gas mass density. 

In order to eliminate as much observational bias as possible, we have defined the redshift 
interval over which to search for DLAs and sub-DLAs with the following criteria. 
As these sightlines were selected due the presence of a galaxy 
within impact parameters of $\rho<150$ kpc, we have excluded $\pm1000$ \kms around the 
redshift of the target galaxy in the search path to avoid biasing 
the sample artificially high  due to the possible clustering of galaxies. 
At $z\ga0.34$ the Mg II \lala 2796, 2803 lines are redshifted into the wavelength coverage of the 
SDSS. This program intentionally excluded sightlines with Mg II absorbers at $z>0.34$ due to the 
possibility of a LLS which would block the FUV flux, which in effect biases the sample against DLA
detection at $z>0.34$. To avoid this bias, we focus only on $z<0.34$.  In the cases where the emission
redshift of the QSO is less than 0.34, we have excluded a 3000 \kms region blueward of the emission
redshift. 

The  sub-DLAs towards 
SDSS J0925+4004, SDSS J0928+6025, and  SDSS J1435+3604 have been excluded from the calculation 
as the absorption redshifts are within 1000 \kms of the redshift of the target galaxy, leaving
3 DLAs and 1 sub-DLA in the final sample. The total path length for the 37 sightlines in this sample using the definition above
is $\Delta z$=11.94, giving d$\mathcal{N}$/$dz=0.25^{+0.24}_{-0.14}$  and 
 d$\mathcal{N}$/dz=0.08$^{+0.19}_{-0.06}$ for DLAs and sub-DLAs respectively, 
and assuming Poisson statistical errors from \citet{Geh86}. The redshift density 
determined here is larger than the value determined 
in \citet{Rao06} of d$\mathcal{N}$/dz=0.079$\pm$0.019, however the large 
statistical errors of this work make this difference unlikely to be statistically significant. 
From a large number of archived HST spectra, \citet{Rib10} find d$\mathcal{N}$/dz=0.37$\pm$0.10
for all absorbers with N(H I)$\ga$17.0. The central value determined here for DLAs alone is 
$\sim$2/3 of the total redshift density of all optically thick systems, again suggesting  that 
the redshift density of DLAs is lower than what is determined here. 

The comoving gas mass density can be determined as
\begin{equation}
\Omega_{\rm H \ I}=\frac{\mu m_H H_0}{c \rho_c}\frac{\Sigma N_{\rm H \ I}}{\Delta X} \label{Eq:Omega}
\end{equation}
where $\mu=1.3$ is the mean molecular mass of the gas taking into account the
contribution of He, $\rho_c$ is the critical  density of the Universe, $H_0$ is the Hubble
constant, and $m_H$ is the mass of the hydrogen atom \citep{WGP05}. The absorption distance dX is defined as
\begin{equation}
\frac{dX}{dz}=\frac{(1+z)^2}{\sqrt{ \Omega_{m}(1+z)^3 + \Omega_{k}(1+z)^2 +   \Omega_\Lambda }}
\end{equation}
and is summed over each sightline in the sample. The total path length for the 37 QSOs in this sample using 
the criteria above is $\Delta$X=15.1.  

We find  \OHI=1.4$^{+1.3}_{-0.7}\times10^{-3}$ 
and  \OHI=4.2$^{+9.6}_{-3.5}\times10^{-5}$ for DLAs and sub-DLAs respectively. 
The neutral gas density in DLAs at $z\sim0.6$ has been determined 
by previous surveys to be \OHI=$(9.7\pm3.6)\times10^{-4}$ 
\citep{Rao00, Rao06}.  The value determined here agrees within the errors
with these previous determinations. The low-$z$ mass density of neutral gas from 
sub-DLA systems has not yet been thoroughly investigated. At $z\sim2.0$, 
\citet{Per05} find \OHI=$(1.9\pm1.3)\times10^{-4}$.

Due to the small number statistics, this sample is dominated by statistical errors. 
Upper and lower bounds on the gas density have been determined by using the upper 
and lower 1$\sigma$ limits on the number of systems, $n^{\pm}$,  from
\citet{Geh86} assuming Poisson statistics. We estimate the 1$\sigma$ range of $\Sigma$ N(H I) in 
Equation \ref{Eq:Omega} as $\Sigma$ N(H I)=$n^{\pm}\langle$N(H I)$\rangle$ where  
$\langle$N(H I)$\rangle$ is the mean \emph{observed} column density in the sample.

\section{Summary and Discussion}

In this work, we have reported on the first Damped Lyman-$\alpha$ systems observed with  the 
Cosmic Origins Spectrograph. We summarize the results of this work as follows. \\

\noindent 1) We have determined the redshift density and gas density
 of DLAs and sub-DLAs in this sample covering a total redshift path of $\Delta z=11.9$ at $z<0.34$. 
We find d$\mathcal{N}$/$dz=0.25^{+0.24}_{-0.14}$  and \OHI=1.4$^{+1.3}_{-0.7}\times10^{-3}$ for DLAs, 
 and for sub-DLAs d$\mathcal{N}$/dz=0.08$^{+0.19}_{-0.06}$ and \OHI=4.2$^{+9.6}_{-3.5}\times10^{-5}$.
 The redshift density of DLAs determined here is several times higher 
than previous measurements of this value at 
$z\sim0.6$ of d$\mathcal{N}$/dz=0.079$\pm$0.019 \citep{Rao00, Rao06}. 
The large statistical errors of our sample make any differences unlikely to be statistically significant. 
No measure of d$\mathcal{N}$/$dz$ 
or \OHI exists for sub-DLAs at $z<1.5$, however the value determined here is lower than d$\mathcal{N}$/dz=0.34$^{+0.33}_{-0.19}$ 
and  \OHI=$(1.9\pm1.3)\times10^{-4}$ seen at $z\sim2.0$ \citep{Per05}. The neutral gas density in DLAs 
measured here is slightly higher than what 
has been found in previous surveys at $z\sim0.6$ which find 
\OHI=$(9.7\pm3.6)\times10^{-4}$  \citep{Rao06}, but the allowed range is fully
 consistent with the previous measurements. Based on 21 cm emission maps of 
galaxies in the nearby Universe, \citep{Zw05} find \OHI=$(3.5\pm0.8)\times10^{-4}$, which is well below
the value determined here.

\noindent 2) We have measured the abundances of several elements of one DLA  in the sample and 
shown that the system has a sub-solar total abundance with [S/H]=$-0.62\pm0.18$. This metallicity is significantly higher
than typical high redshift DLAs, which have average metallicities $\langle$[Zn/H]$\rangle\sim-1.25$. 
The kinematical structure of the DLA is remarkably simple. Only two components are needed to fit the profiles and the 
velocity width is well below that of the average DLA population with \dv=52 \kms. Remarkably, the
kinematics are similar to those of the extremely metal poor DLAs from \citet{Pet08}, even though
the metallicity of this DLA is $\sim$100 times greater than the DLAs in that sample. \\

\noindent 3) Followup imaging with WFC3 reveals several star forming and presumably gas rich galaxies within 
10$\arcsec$ of the QSO, as well as a large spiral galaxy at small impact parameter to the bright 
QSO. At first glance, any of these candidates appear to be  promising candidates for the host galaxy of the DLA
system. Followup spectroscopy of the galaxies reveal that none of them are at a similar redshift to the DLA.
Spectra of an H II region in the spiral galaxy show strong narrow emission
lines at the redshift of the QSO. This, combined with the close alignment 
of the nucleus and disk of the galaxy  leads us to identify the spiral galaxy
as the host of the QSO. The QSO host is luminous, with M$_{F625W}=-24.4$
based on a decomposition of the WFC3 image with \textsc{galfit}. 
A small feature is seen $\sim$0.7
arcseconds from  the center of the QSO nucleus after PSF subtraction, adding 
another host galaxy candidate  of the DLA. Even with this supporting data, the origin of the DLA host 
galaxy remains enigmatic and exemplifies some of the difficulties of discovering the host galaxies of such systems. \\

\noindent 4) Independent of the morphological properties of the DLA host galaxy, we 
conclude that the sightline likely passes through the warm neutral medium of the host as 
the metals are very mildly depleted with [S/Fe]=$+0.24\pm0.22$, [S/Ni]=$+0.35\pm0.22$ 
and [S/Ti]=$+0.28\pm0.15$ and both C II$^{*}$ and C I are not detected. 
The depletion levels and low cold gas content are characteristic
of higher $z$ DLA systems \citep{Pro02, Mei06, Not09a}. We find that the line of sight depletion factor
F$_{*}=-0.11\pm0.24$, characteristic of sightlines through the halo of the Milky Way, reaffirming the 
conclusion that the sightline is passing through warm neutral gas.    \\

In a forthcoming paper we will report on the metallicities and depletions in the other
DLAs and sub-DLAs in this sample, as well as the N(H I)-weighted mean metallicity at $z<0.34$. 
Few observations of S or Zn, undepleted elements in the ISM and 
accurate metallicity indicators, exist at $z<0.5$ which spans $\sim40$ percent of the age of the universe. 
No observations of Zn or S in sub-DLA systems exist in this redshift
range. At low redshifts, where there has been more time for the successive buildup of metals in the ISM of galaxies
and the abundances of DLAs should be higher than at high redshift, it will be possible to see if DLAs show 
an increase in the [N/$\alpha$] ratio characteristic of secondary production of N. 
Our data show that over its lifetime, the Cosmic Origins Spectrograph will have the ability to measure the abundances and
physical properties of QSO absorbers at $z\sim0$ with nearly the same precision as what can be done 
at $z\sim2$ with large ground based observatories.

\section*{Acknowledgments}
Some of the data presented herein were obtained at the W.M. Keck Observatory, 
which is operated as a scientific partnership among the California Institute of Technology, 
the University of California and the National Aeronautics and Space Administration. The Observatory was 
made possible by the generous financial support of the W.M. Keck Foundation. 
The authors wish to extend special thanks to those of Hawaiian ancestry on whose sacred mountain 
we are privileged to be guests. Without their generous hospitality these observations would not have been 
possible. Financial support for this research was provided by NASA grants HST-GO-11598.03-A and NNX08AJ44G. 
 J.X.P. also acknowledges support from NSF grant (AST-0709235).


\begin{thebibliography}{99}
\bibitem[Bahcall et al.(1997)]{Bah97} Bahcall J.N., Kirhakos S., Saxe D.H., Schneider D.P., 1997, ApJ, 479, 642
\bibitem[Becker, White $\&$ Helfand(1995)]{BWH95} Becker R.H., White R.L., Helfand D.J., 1995, ApJ, 450, 559
\bibitem[Blanton et al.(2003)]{Blan03} Blanton M.R., Hogg D.W., Bahcall N.A., Brinkman J., Britton M., Connolly A.J., Csabai I., Loveday J., et al., 2003, ApJ, 592, 819
\bibitem[Borthakur et al.(2010)]{Bor10} Borthakur S., Tripp T.M., Yun M.S., Momijian E., Meiring J.D., Bowen D.V., York D.G., 2010, ApJ, 713, 131 
\bibitem[Bowen, Tripp $\&$ Jenkins(2001)]{Bow01} Bowen D.V., Tripp T.M., Jenkins E.B., 2001, AJ, 121 1456 
\bibitem[Bowen et al.(2005)]{Bow05} Bowen D.V., Jenkins E.B., Pettini M., Tripp T.M., 2005, ApJ, 635, 880
\bibitem[Centuri\'on et al.(2003)]{Cent03} Centuri\'on M., Molaro P., Vladilo G., P\'eroux C., Levshakov S.A., D'Odorico V., 2003, A$\&$A, 403, 55
\bibitem[Elmegreen $\&$ Parravano(1994)]{Elm94} Elmegreen B.G., Parravano A., 1994, ApJ, 435, 121 
\bibitem[Fitzpatrick $\&$ Spitzer(1994)]{Fitz94} Fitzpatrick E. L., Spitzer, L., 1994, ApJ, 427, 232
\bibitem[Froning $\&$ Green(2009)]{Fro09} Froning C.S., Green J.C., Ap$\&$SS, 320, 181
\bibitem[Gehrels(1986)]{Geh86} Gehrels N., 1986, ApJ, 303, 336
\bibitem[Ghavamian et al.(2009)]{Ghav09} Ghavamian, P., et al. 2009, Preliminary Characterization of the Post-Launch Line Spread Function of COS, \texttt{http://www.stsci.edu/hst/cos/documents/isrs/}
\bibitem[Goddard, Kennicutt $\&$ Weber(2010)]{Goddard10} Goddard Q.E., Kennicutt R.C., Ryan-Weber E.V., 2010, MNRAS, 405, 2791
\bibitem[Haardt $\&$ Madau(1996)]{HM96}Haardt F., Madau P., 1996, ApJ, 461, 20
\bibitem[Henry $\&$ Prochaska(2007)]{Hen07} Henry R. B. C., Prochaska J.X., 2007, PASP, 119, 962
\bibitem[Hamilton, Casertano $\&$ Turnshek(2002)]{Ham02} Hamilton T.S., Casertano S., Turnshek D.A., 2002, ApJ, 576, 61
\bibitem[Jenkins $\&$ Tripp(2001)]{JT01} Jenkins E.B., Tripp T.M., 2001, ApJS, 137, 297
\bibitem[Jenkins $\&$ Tripp(2006)]{JT06} Jenkins E.B., Tripp T.M., 2006, ApJ, 637, 548
\bibitem[Jenkins(2009)]{Jen09} Jenkins E.B., 2009, ApJ, 700, 1299
\bibitem[Kennicutt(1998)]{Ken98} Kennicutt R.C., 1998, ApJ, 498, 541
\bibitem[Kulkarni et al.(2010)]{Kul10}Kulkarni V.P.,  Khare P., Som D., P\'eroux C., York D.G., Meiring J.D., Lauroesch J.T., 2010, New Astronomy, in press
\bibitem[Liszt(2002)]{Liszt02} Liszt H., 2002, A$\&$A, 389, 343
\bibitem[Ledoux et al.(2006)]{Led06} Ledoux, C. Petitjean, P., Moller, P., Fynbo, J., Srianand R., 2006, A$\&$A, 457, 71
\bibitem[Lehner, Wakker $\&$  Savage(2004)]{Leh04} Lehner N.,  Wakker B.P., Savage B.D., 2004, ApJ, 615, 757
\bibitem[Lodders(2003)]{Lodd03} Lodders, K., 2003, ApJ, 591, 1220
\bibitem[Lupton et al.(2004)]{Lup04} Lupton R., Blanton M.R., Fekete G., Hogg D.W., O'Mullane W., Szalay A., Wherry N., 2004, PASP, 116,133
\bibitem[Meiring et al.(2006)]{Mei06} Meiring J.D., Kulkarni V.P., Khare P., Bechtold J., York D.G., Cui J., Lauroesch J.T., Crotts A.P.S.,Nakamura O., 2006, MNRAS,370, 43
\bibitem[Meiring et al.(2009A)]{Mei09a} Meiring J.D., Kulkarni V.P., Lauroesch J.T., P\'eroux C., Khare P.,  York D.G., 2009, MNRAS, 393, 1513
\bibitem[Meiring et al.(2009B)]{Mei09b} Meiring J.D.,  Lauroesch J.T., Kulkarni V.P., P\'eroux C., Khare P.,  York D.G., 2009, MNRAS, 397, 2037
\bibitem[Marigo(2001)]{Mar01} Marigo P., 2001, A$\&$A, 370, 194
\bibitem[Noterdaeme et al.(2009)]{Not09a} Noterdaeme P., Ledoux C., Petitjean P., Srianand R., 2009, A$\&$A, 327, 336
\bibitem[Noterdaeme et al.(2009)]{Not09b} Noterdaeme P., Petitjean P., Ledoux C., Srianand R., 2009, A$\&$A, 1087, 1098
\bibitem[Noterdaeme, Srianand $\&$ Mohan(2010)]{Not10} Noterdaeme P., Srianand R., Mohan V., 2010, MNRAS, 403, 906
\bibitem[Peng et al.(2010)]{Peng10} Peng C.Y., Ho L.C., Impey C.D., Rix H-W., 2010, AJ, 139, 2097
\bibitem[Pettini et al.(2008)]{Pet08} Pettini M., Zych B.J., Steidel C.C., Chaffee F.H., 2008, MNRAS, 385, 2011
\bibitem[Prochaska $\&$ Wolfe(1997)]{PW97} Prochaska J. X.,  Wolfe A. M., 1997, ApJ, 487, 73
\bibitem[Prochaska $\&$ Wolfe(2002)]{PW02} Prochaska J.X., Wolfe A.M., 2002, ApJ, 566, 68
\bibitem[Prochaska et al.(2003)]{Pro03} Prochaska J.X., Gawiser E., Wolfe A.M., Castro S., Djorgovski S. G., 2003, ApJ, 595, L9
\bibitem[Prochaska et al.(2002)]{Pro02} Prochaska J.X., Howk J.C., O'Meara J.M., Tytler D., Wolfe A.M., Kirkman D., Lubin D., Suzuki N., 2002, ApJ, 571, 693
\bibitem[Prochaska, Herbert-Fort $\&$ Wolfe(2005)]{Pro05} Prochaska J.X., Herbert-Fort S., Wolfe A.M., 2005, ApJ, 635, 123
\bibitem[Prochaska et al.(2006)]{Pro06} Prochaska J.X., O'Meara J.M., Herbert-Fort S., Burles S., Prochter G.E., Bernstein R.A., 2006, ApJL, 648, 97
\bibitem[P\'eroux et al.(2003a)]{Per03a} P\'eroux C., Dessauges-Zavadsky M., D'Odorico S., Kim T.S., McMahon R., 2003, MNRAS, 345, 480
\bibitem[P\'eroux et al.(2003b)]{Per03b} P\'eroux C., McMahon R.G., Storrie-Lombardi L.J., Irwin M.J., 2003, MNRAS, 346, 1103
\bibitem[P\'eroux et al.(2005)]{Per05} P\'eroux C., Dessauges-Zavadsky  M.,  D'Odorico S.,  Kim T.S.,  McMahon R.G., MNRAS, 363, 479
\bibitem[Rao $\&$ Turnshek(2000)]{Rao00} Rao S.M., Turnshek D.A., 2000, ApJ, 130, 1
\bibitem[Rao, Turnshek $\&$ Nestor(2006)]{Rao06} Rao S.M., Turnshek D.A., Nestor D.B., 2006, ApJ, 636, 610
\bibitem[Ribaudo, Lehner $\&$ Howk(2010)]{Rib10} Ribaudo J., Lehner N., Howk C.J., 2010, ApJ, submitted
\bibitem[Savage $\&$ Sembach(1996)]{SS96} Savage B.D., Sembach K.R., 1996, ARAA, 34, 279
\bibitem[Schaye(2004)]{Sch04} Schaye J., 2004, ApJ, 609, 667
\bibitem[Schlegel, Finkbeiner $\&$ Davis(1998)]{Sch98} Schlegel D.J., Finkbeiner D.P., Davis M., 1998, ApJ, 500, 525
\bibitem[Sembach $\&$ Savage(1996)]{Sem96} Sembach K.R., Savage B.D., 1996, ApJ, 457, 211
\bibitem[Srianand et al.(2005)]{Sri05} Srianand R., Petitjean P., Ledoux C., Ferland G., Shaw G., 2005, MNRAS, 362, 549
\bibitem[Storrie-Lombardi $\&$ Wolfe(2000)]{SL00} Storrie-Lombardi L.J., Wolfe A.M., 2000, ApJ, 543, 552
\bibitem[Thom et al.(2010)]{Thom10} Thom C., Werk J.E., Tumlinson J., Prochaska J.X., Meiring J.D., Tripp T.M., Sembach K.R., 2011, ApJ, submitted
\bibitem[Toomre(1964)]{Toom64} Toomre A., 1964, ApJ, 139, 1217
\bibitem[Tremonti et al.(2004)]{Trem04} Tremonti C.A., Heckman T.M., Kauffmann G., Brinchmann J., Charlot S., White S.D.M., Seibert M., Peng E.W., et al. 2004, ApJ, 613, 898
\bibitem[Tripp et al.(2005)]{Tripp05}  Tripp T.M.,  Jenkins E.B., Bowen D.V., Prochaska J.X., Aracil B., Ganguly R., 2005, ApJ, 619, 714
\bibitem[Tumlinson et al.(2010)]{Tum10} Tumlinson J.T., Werk J.K., Thom C.T., Meiring J.D., Prochaska J.X., Tripp T.M., Okrochkov M., Sembach K.R., 2011, submitted
\bibitem[Van Zee et al.(1998)]{Zee98} Van Zee L., Salzer J.J., Haynes M.P., O'Donoghue A.A., 1998, ApJ, 116, 2805
\bibitem[Van Zee $\&$ Haynes(2006)]{Zee06} Van Zee L., Haynes M.P., 2006, ApJ, 636, 214
\bibitem[Vogt et al.(1994)]{Vogt94} Vogt et al., 1994, Proc. SPIE, 2198, 362
\bibitem[Morton(2003)]{Mort03} Morton D.C., 2003, ApJS, 149, 205
\bibitem[Wakker(2006)]{Wak06} Wakker B.P., 2006, ApJ, 163, 282
\bibitem[Welty et al.(1997)]{Wel97} Welty D. E., Lauroesch J.T., Blades J.C., Hobbs L.M., York, D.G., 1997, ApJ, 489, 672
\bibitem[Wolfe et al.(1995)]{Wol95} Wolfe A.M., Lanzetta K.M., Foltz C.B., Chaffe F., 1995, ApJ, 454, 698 
\bibitem[Wolfe, Prochaska $\&$ Gawiser(2003)]{WPG03} Wolfe A.M., Prochaska J.X., Gawiser E., 2003, ApJ, 593, 215
\bibitem[Wolfe, Gawiser $\&$ Prochaska(2005)]{WGP05} Wolfe A.M., Gawiser E., Prochaska J.X., 2005, Annu. Rev. Astron. Astrophys., 43, 861
\bibitem[Wright(1991)]{Wright91} Wright E.L., 1991, ApJ, 381, 200
\bibitem[York et al.(2000)]{York00} York D.G., Adelman J., Anderson J.E., Jr., Anderson S.F., Annis J., Bahcall N.A., Bakken J. A., Barkhouser R.,  et al. 2000, AJ, 120, 1579
\bibitem[Zwaan et al.(2005)]{Zw05} Zwaan M. A., Meyer M.J., Staveley-Smith L., Webster R.L., 2005, MNRAS, 359, 30
\end{thebibliography}
\end{document}